\documentclass{INTERSPEECH2023}

\usepackage{booktabs, multirow, graphicx, subcaption, threeparttable, etoolbox, nccmath}

\AtBeginEnvironment{tabular}{\fontsize{9}{11}\selectfont}

\interspeechcameraready

\title{Outlier-aware Inlier Modeling and Multi-scale Scoring for Anomalous Sound Detection via Multitask Learning}
\name{Yucong Zhang$^1$, Hongbin Suo$^2$, Yulong Wan$^2$, Ming Li$^{1\ast}$\thanks{$\ast$ Corresponding author: Ming Li.}}
\address{
$^1$Data Science Research Center, Duke Kunshan University, Kunshan, China \\
$^2$Data \& AI Engineering System, OPPO, Beijing, China
}
\email{ming.li369@dukekunshan.edu.cn}

\begin{document}

\maketitle
 
\begin{abstract}
This paper proposes an approach for anomalous sound detection that incorporates outlier exposure and inlier modeling within a unified framework by multitask learning. While outlier exposure-based methods can extract features efficiently, it is not robust. Inlier modeling is good at generating robust features, but the features are not very effective. Recently, serial approaches are proposed to combine these two methods, but it still requires a separate training step for normal data modeling. To overcome these limitations, we use multitask learning to train a conformer-based encoder for outlier-aware inlier modeling. Moreover, our approach provides multi-scale scores for detecting anomalies. Experimental results on the MIMII and DCASE 2020 task 2 datasets show that our approach outperforms state-of-the-art single-model systems and achieves comparable results with top-ranked multi-system ensembles.
\end{abstract}


\section{Introduction}
Anomalous sound detection (ASD) is a crucial technology for identifying anomalous sounds in various industries~\cite{kamat2020anomaly, tanuska2021smart}. It helps detect and isolate sound anomalies that may indicate malfunctions or potential dangers. The importance of ASD lies in its ability to prevent accidents and improve operational efficiency by detecting and addressing issues before they become critical. With the advent of smart technologies and the Internet of Things~(IoT), the demand for accurate anomalous sound detection solutions continues to grow~\cite{wu2021network}, making it a critical technology in today's industrial landscape.

Over the past two years, the two dominant approaches for ASD have been Inlier Modeling~(IM) and Outlier Exposure~(OE)~\cite{koizumi2020description, kawaguchi2021description}. Given that it is often more challenging to obtain anomalous data compared to normal data~\cite{chandola2009anomaly}, unsupervised methods that do not require anomalous data are frequently used for the task. IM is such method that involves modeling the probability distribution of normal data. Well-known techniques such as AutoEncoders~(AE)~\cite{giri2020groupMADE, daniluk2020IDCAE, hayashi2020conformer, suefusa2020IDNN}, Local Outlier Factor~(LOF)~\cite{breunig2000lof}, Gaussian Mixture Models~(GMM)~\cite{scott2004outlier, liu2019outlier}, and Normalizing Flows~(NF)~\cite{dohi2021flow} have been explored within the scope. However, IM is hard to extract effective features~\cite{kuroyanagi2022improvement}.

Although it is hard to collect anomalous data, pseudo-anomalous data can be generated to compensate for this shortfall, leading to the development of OE-based methods. These methods concentrate on learning the outlying decision boundaries of normal data by classifying normal and pseudo-anomalous data. OE-based methods are surprisingly effective at identifying useful features. In DCASE 2020 Task 2~\cite{koizumi2020description}, several top-performing teams employed OE-based techniques in their system and demonstrated their effectiveness~\cite{giri2020groupMADE,primus2020reframing,inoue2020detection}. Nonetheless, OE-based methods can be unreliable and underperform when the normal and pseudo-anomalous data are either too similar or too distinct~\cite{koizumi2020description,kawaguchi2021description,dohi2021flow}.

Recently, to address the challenges posed by IM and OE, two hybrid approaches have been proposed and have demonstrated great success in ASD~\cite{kawaguchi2021description}. These hybrid approaches are referred to as the parallel approach and the serial approach. The parallel approach involves combining anomaly scores from both IM and OE to make up for each other's weaknesses~\cite{giri2020groupMADE, lopez2021ensemble}. However, this requires multiple models with different training processes, which increases the cost and difficulty of development and maintenance. In contrast, the serial approach involves using IM and OE in a sequential manner~\cite{kuroyanagi2022improvement, morita2021anomalous, wilkinghoff2021utilizing}. It first uses OE-based method to train an encoder, then utilizing IM-based method to train a normal data distribution fitting the embeddings extracted by the encoder. Although it avoids parallel training for multiple models, it still needs two training processes to form the normal data distribution for future anomaly scoring.

Multitask learning~(MTL) has been widely discussed in anomalous video detection~\cite{georgescu2021anomaly, doshi2022multi}. They use MTL to better capture motion patterns that traditional methods might ignore. MTL can also be employed in ASD to enrich the signal features that are extracted by OE~\cite{kuroyanagi2021anomalous, kuroyanagi2022improvement}. In~\cite{kuroyanagi2022improvement}, the authors suggest that the performance of the serial approach can be improved by training OE with MTL. Enlightened by their work, we train a conformer-based encoder with MTL to learn inlier modeling with the awareness of outlying decision boundaries. The key highlights of our work are as follows:
\begin{enumerate}[leftmargin=.35cm]
    \item Our approach comprises only one model and yet takes into consideration the concept of both IM and OE, making it simple to develop and maintain.
    \item During training, multitask learning enables the encoder to learn inlier properties within the outlying decision boundaries by considering both the outlier and inlier data.
    \item During inference, our approach can provide comprehensive scores by considering three different aspects, which enables us to detect anomalies in a more comprehensive way.
    \item The experimental results on both DCASE 2020 dataset and MIMII dataset show the state-of-the-art performance and demonstrate the robustness of our approach under different levels of noisy conditions.
\end{enumerate}

\begin{figure*}[!htbp]
    \centering
    \includegraphics[width=.85\textwidth]{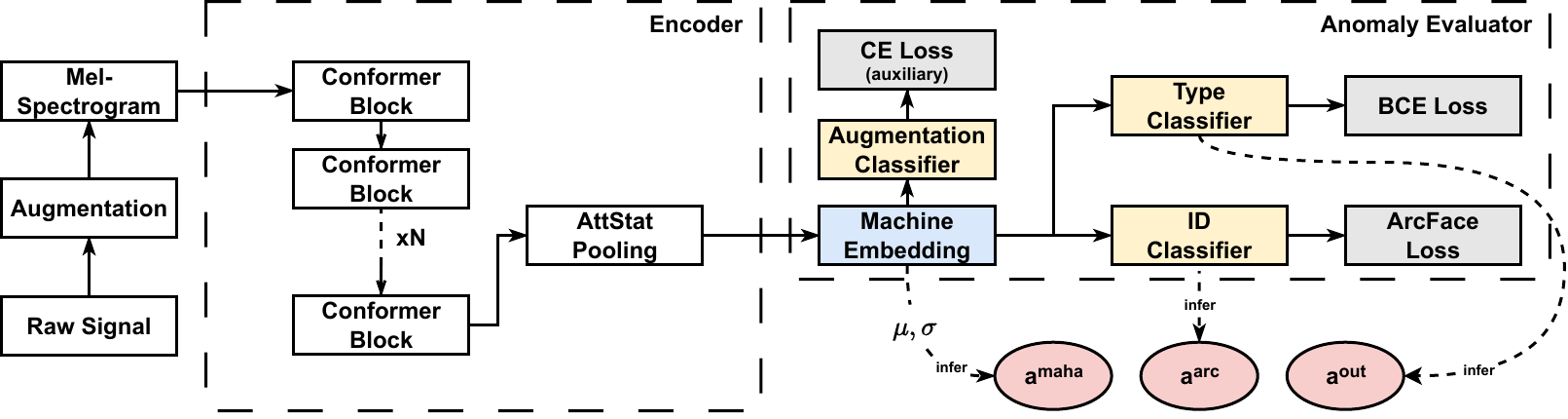}
    \vspace{-1mm}
    \caption{The overview of the proposed approach}
    \label{fig:overview}
    \vspace{-6mm}
\end{figure*}

\section{Proposed Method}
Figure~\ref{fig:overview} provides an overview of our proposed approach. The whole learning scheme is comprised of a front-end encoder that generates embeddings and a back-end anomaly evaluator consisting of three classifiers for different tasks. In Section~\ref{subsec:encoder}, we introduce a conformer-based encoder. In Section~\ref{subsec:MTL}, we explain different tasks for MTL. Then, we describe how we train the model using MTL in the training phase. Finally, we provide detailed explanation on how to compute multi-scale anomaly scores during inference with our approach in Section~\ref{subsec:inference}.

\vspace{-1mm}\subsection{Conformer-based Encoder}
\label{subsec:encoder}
Our front-end encoder is made up of the conformer modules described in~\cite{gulati2020conformer}. Each conformer module includes two feed-forward networks~(FFN) that sandwich a multi-head self-attention~(MHSA) module followed by a convolution~(Conv) module. 
Since the conformer module contains both the attention mechanism and convolution, it can effectively capture both local information and long-range dependencies from the spectral feature. Local information is crucial in differentiating between inlier differences among machines with similar sounds, while long-range information is essential for capturing overall characteristics that help to define the boundaries of the learnt representations. In combination, these two types of information provide a comprehensive understanding of the spectral features.

\vspace{-1mm}\subsection{Multitask Learning}
\label{subsec:MTL}
To effectively capture both inlier and outlying attributes, our method employs MTL with three different losses. The first task is to learn inlier properties by distinguishing among different machine IDs with same machine type using Additive Angular Margin Loss~(ArcFace)~\cite{deng2019arcface}. The second task is to identify outlying decision boundaries by determining whether the current signal belongs to the target machine type. The final task is to enable robust training by identifying different types of augmentation applied to the original signal.

\vspace{-2mm}\subsubsection{Task 1: Learn inlier properties}
\label{subsubsec:IM}
The first objective of  the anomaly evaluator is to learn the inherent properties of normal data. To model the normal data distribution, we first build a classifier~($\mathcal{C}_{\text{id}}$) that identifies different machine IDs with the same machine type using ArcFace. Then, we compute the mean and covariance of the deep features for each machine ID using the normal data, which are important statistics describing the data distribution. We choose a classification-based method to explore inlier characteristics because similar methods have already demonstrated their efficiency in identifying useful features in OE-based approaches~\cite{kawaguchi2021description}.

ArcFace projects the softmax function into the angular space, providing geometric interpretations for the model. Suppose we have a linear classifier that can distinguish among K different machine IDs, we can compute the output probability for the $i^{th}$ sample $\mathbf{x}_i$ as  $\mathbf{y}_{i}^{\prime}=[\mathbf{y}_{i,1}^{\prime}, \dots, \mathbf{y}_{i,K}^{\prime}]^{\mathrm{T}}$. The probability of the $k^{th}$ class $\mathbf{y}_{i,k}^{\prime}$ is computed as $\mathrm{\mathbf{W}}_{k}^{\mathrm{T}}\mathbf{x}_{i}+\mathbf{b}_k$, where $\mathbf{W}_{k}$ is the $k^{th}$ column of the weight matrix $\mathbf{W}$ and $\mathbf{b}_k$ represents the bias term. We refer to $\mathbf{W}_{k}$ as the $k^{th}$ ArcFace anchor. Assuming we have $||\mathbf{W}_{k}||=1$ and $\mathbf{b}_k=0$ for all $k$, we can rewrite the original equation as $\mathbf{y}_{i,k}^{\prime}=||\mathbf{x}_{i}||\cos{\mathbf{\theta}_{i,k}}$, where $\mathbf{\theta}_{i,k}$ is the angle between the input $\mathbf{x}_{i}$ and the $k^{th}$ anchor. Therefore, the loss $\mathcal{L}_{\text{id}}$ is derived as follows:
\begin{equation}
\label{eq:arcface_loss}
\begin{aligned}
\mathcal{L}_{\text{id}} &= -\sum_{i=1}^{N} \left[ \log \frac{\exp(\mathbf{a}_{i, \mathrm{y}_i})}{\exp(\mathbf{a}_{i, \mathbf{y}_i})+\sum_{j=1, j \neq \mathbf{y}_i}^K \exp(\mathbf{a}_{i,j}^\prime)} \right] \\
\mathbf{a}_{i,k} &= s \cos\left({\mathbf{\theta}_{i,k}}+m\right),\quad \mathbf{a}_{i,k}^\prime = s \cos\left({\mathbf{\theta}_{i,k}}\right) \\
\mathbf{a}_{i} &= [\mathbf{a}_{i,1},\dots,\mathbf{a}_{i,k}^\prime,\dots,\mathbf{a}_{i,K}]^{\mathrm{T}}, 1\leq k \leq K
\end{aligned}
\end{equation}
where $\mathrm{\mathbf{a}}_{i}=\mathcal{C}_{\text{id}}(\mathbf{x}_{i})$, each elements $\mathbf{a}_{i,k} \in \mathrm{\mathbf{a}_i}$ denotes the probability that the $i^{th}$ sample belongs to the $k^{th}$ class. $s$ and $m$ are hyper-parameters that control radius and margin respectively.

\begin{figure}[!htbp]
\vspace{-2mm}
\centering
\begin{minipage}[b]{0.23\textwidth}
\includegraphics[width=\textwidth]{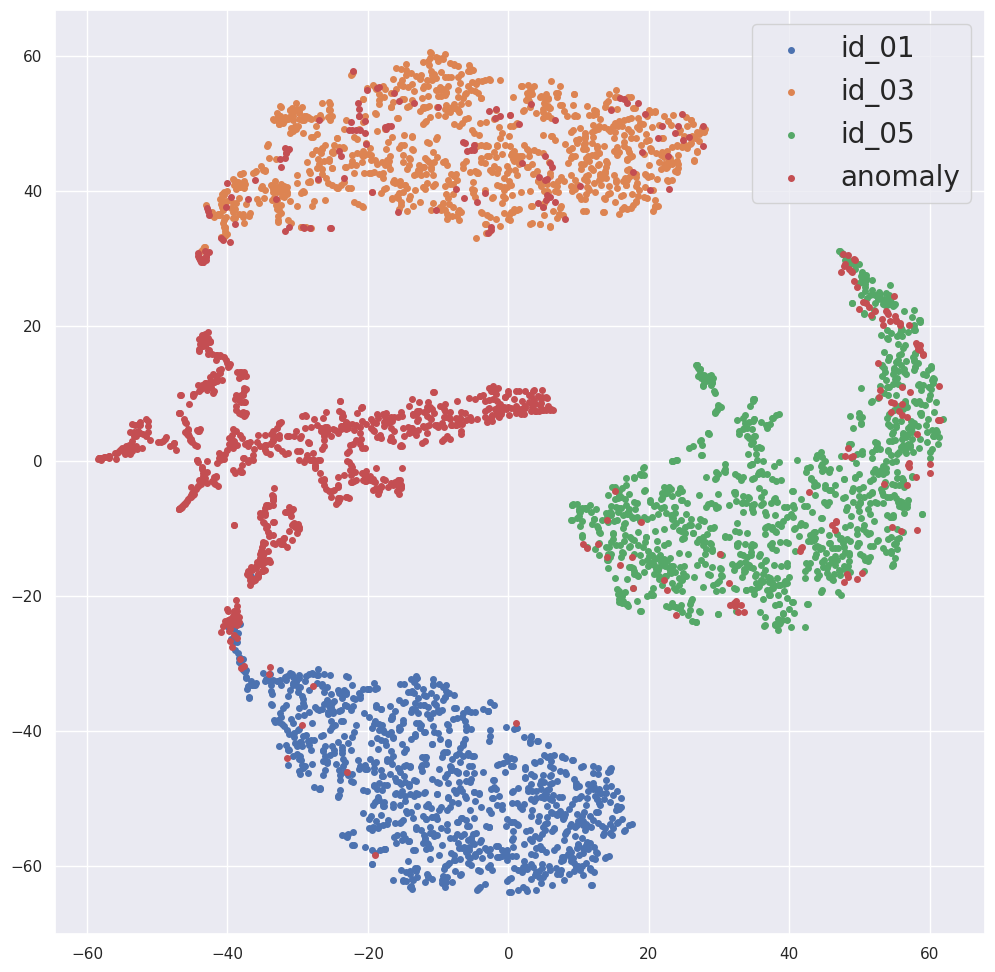}
\subcaption{Fan}
\end{minipage}
\hfill
\begin{minipage}[b]{0.23\textwidth}
\includegraphics[width=\textwidth]{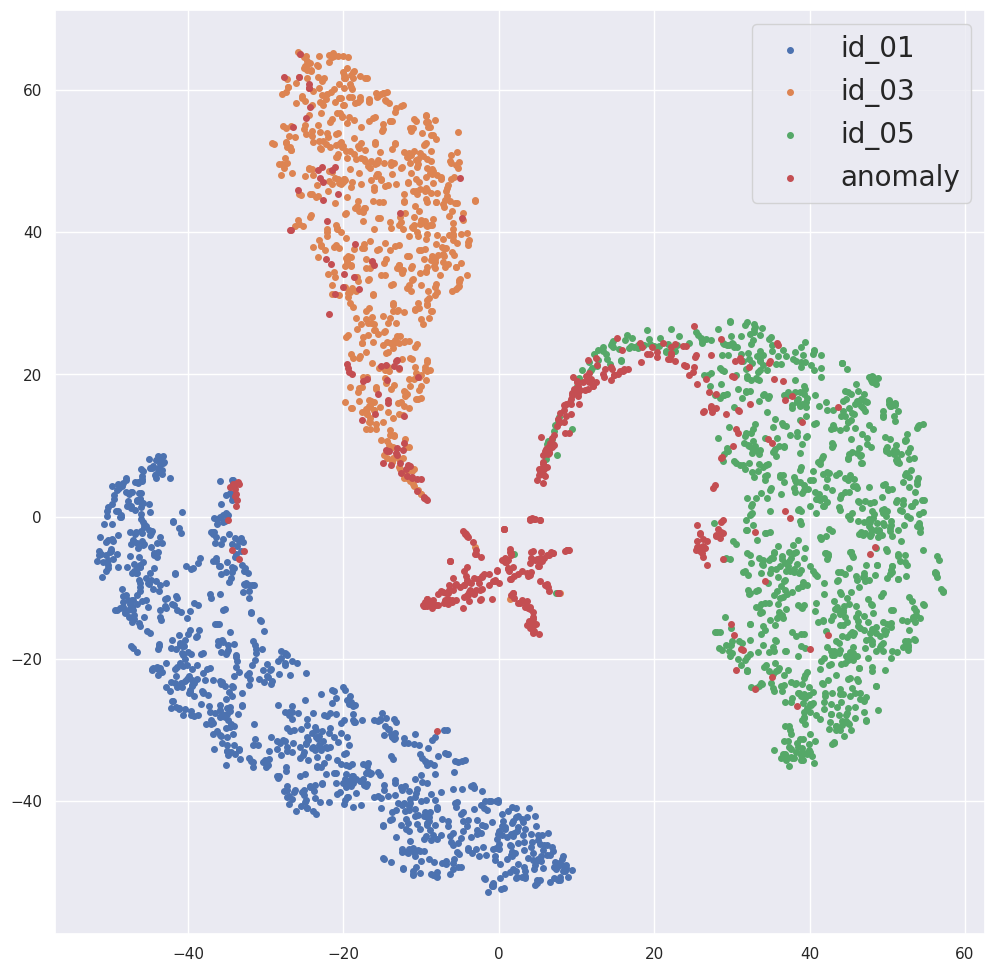}
\subcaption{Pump}
\end{minipage}
\vfill
\begin{minipage}[b]{0.23\textwidth}
\includegraphics[width=\textwidth]{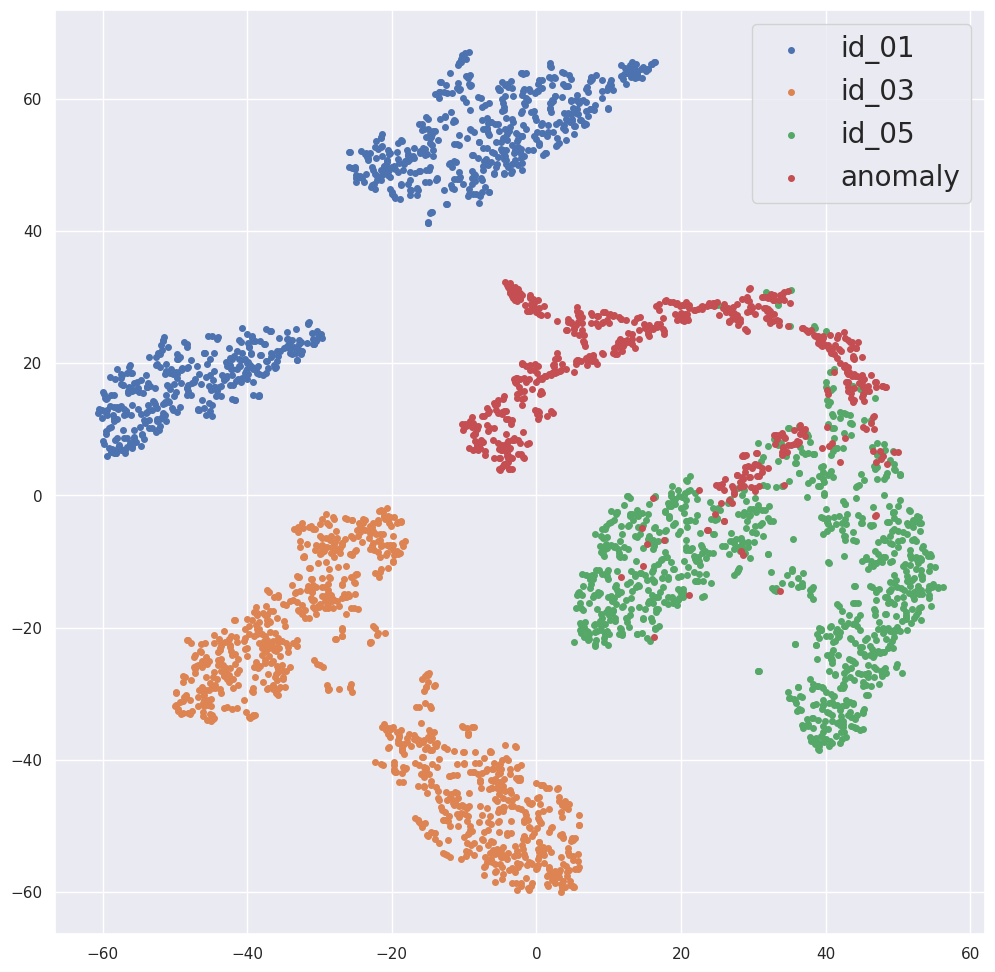}
\subcaption{Valve}
\end{minipage}
\hfill
\begin{minipage}[b]{0.23\textwidth}
\includegraphics[width=\textwidth]{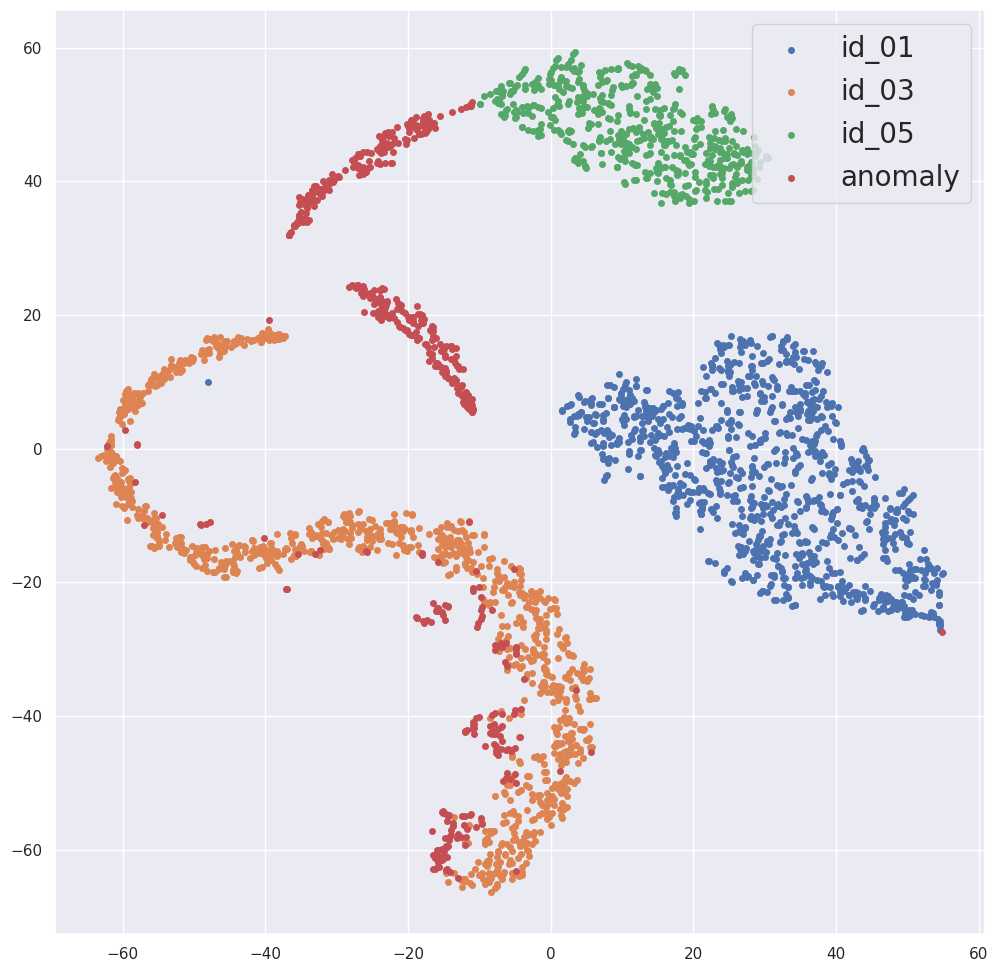}
\subcaption{Slider}
\end{minipage}
\caption{Visualization of data distribution after dimension reduction using t-SNE on DCASE~2020 Evaluation Dataset}
\label{fig:latent_space}
\vspace{-2mm}
\end{figure}

From a geometric view~(shown in Figure~\ref{fig:latent_space}), the ArcFace loss attempts to group together features with the same machine ID by minimizing the angles between them while pushing away those with different IDs. As a result, the distribution of the features is naturally formed during the training process, the inlier properties such as mean and variance can then be calculated.

\vspace{-2mm}\subsubsection{Task 2: Find outlying decision boundaries}
\label{subsubsec:OE}
Our second task aims at identifying decision boundaries of the normal data. Inspired by OE-based methods, we treat signals of the target machine type as normal data and signals of other types as pseudo-anomalous data. Then, we can use a binary classifier~($\mathcal{C}_{\text{type}}$) to distinguish between normal and pseudo-anomalous data. To enable training, binary cross-entropy~(BCE) is picked as the loss function. Suppose $\mathbf{x}_{i} \in \mathbb{R}^{\mathrm{d}}$ is the $i^{th}$ deep feature derived from the conformer-based encoder and $\mathrm{d}$ denotes the dimension, then we have the following:
\begin{equation}
    \mathcal{L}_{\text{type}} = -\sum_{i=1}^{N} \left[\mathbf{y}_i\cdot\log(\mathbf{a}_i) + (1-\mathbf{y}_i)\cdot\log(1-\mathbf{a}_i)\right]
\end{equation}
where $\mathbf{a}_i = \mathcal{C}_{\text{type}}(\mathbf{x}_i)$ and $\mathbf{y}_{i}$ denotes the $i^{th}$ label. If $\mathbf{x}_i$ belongs to the target type, $\mathbf{y}_{i}$ is 1, otherwise it is 0.

\vspace{-2mm}\subsubsection{Task 3~(Auxiliary): Enhance robustness}
\label{subsubsec:robust}
Our third objective is to improve the model's ability to recognize key features and prevent overfitting, which we accomplish through two processes. First, we augment the original data by applying various types of operations as described in~\cite{hojjati2022self}, such as pitch/time shifting, time stretching, fading in/out, white noise injection, and time/frequency masking. This enables us to introduce more variations to the original data, compelling the model to capture crucial parts of the features. Second, in case false alarms might be caused by the variations introduced by the augmentation, we develop an auxiliary classifier~($\mathcal{C}_{\text{aug}}$) followed by cross-entropy loss to assist training. This classifier identifies the type of augmentation applied to the original signal. The loss $\mathcal{L}_{\text{aug}}$ has the same equation as Equation~\ref{eq:arcface_loss}, except that $\mathbf{a}_{i,k}$ and $\mathbf{a}_{i,k}^\prime$ are the output of the linear classifier~$\mathcal{C}_{\text{aug}}$ with the same value.

The overall loss of our proposed method is calculated by the weighted sum of all the three losses:
\begin{equation}
\label{eq:overall_loss}
\mathcal{L}=\mathcal{L}_{\text{type}}+\alpha \mathcal{L}_{\text{id}}+\beta \mathcal{L}_{\text{aug}}
\end{equation}
where $\alpha,\beta$ are the hyper-parameters controlling the weights. In our work, we let $\alpha=\beta=1$. 

\vspace{-2mm}\subsubsection{Training strategy}
\label{subsubsec:strategy}
OE-based methods are unstable and easily get overfitting if the pseudo-anomalous data are too distinct or too similar with the normal data~\cite{koizumi2020description,kawaguchi2021description,dohi2021flow}. Hence, to tackle this issue, we adopt a two-stage inside-out training strategy. First, we freeze $\mathcal{C}_{\text{type}}$ and train the encoder by $\mathcal{C}_{\text{id}}$ using the normal data that only comes from the target machine type. This allows us to pre-train the encoder focusing on the inlier properties. Then, with the initial weights, we unfreeze $\mathcal{C}_{\text{type}}$ and add pseudo-anomalous data to further train the encoder for several epochs. Different from traditional IE-based methods, our approach considers outlier data when modeling normal data distribution, leading to a more effective feature extractor.

\vspace{-1mm}\subsection{Inference}
\label{subsec:inference}
Multitask learning allows the anomaly evaluator to take into account multiple perspectives during the model training. As a result, three different anomaly scores are generated in the inference phase: $\mathrm{a^{\text{out}}}$, $\mathrm{a^{\text{arc}}}$ and $\mathrm{a^{\text{maha}}}$.

The outputs of $\mathcal{C}_{\text{type}}$ and $\mathcal{C}_{\text{id}}$ provide distinct perspectives for the anomaly evaluator when scoring anomalies. The likelihood output $\mathrm{a^{\text{out}}}$ from $\mathcal{C}_{\text{type}}$ reflects the effectiveness of the outlying decision boundaries in determining if the signal belongs to the target machine type. Meanwhile, the probability output $\mathrm{a^{\text{arc}}}$ from $\mathcal{C}_{\text{id}}$ reveals how likely the signal belongs to a particular machine ID. In Section~\ref{subsubsec:IM}, we refer to the column vectors of the weight matrix in ArcFace as ArcFace anchors. 
If the deep feature is closer to the target anchor, the likelihood of that feature being normal is higher. Therefore, $\mathrm{a^{\text{arc}}}$ can also serve as a metric of deviation from the normal data distribution.

The last score, $\mathrm{a}^{\text{maha}}$, is derived by calculating the Mahalanobis distance between the test data and the normal data distribution. After training, the encoder is used to extract deep features for all the normal data and categorize them by machine type and IDs. Suppose $\boldsymbol{\mu}_{t,i}$ and $\boldsymbol{\Sigma}_{t,i}$ represents the mean and covariance of the normal features of machine type $t$ and ID $i$, and $\mathbf{x}$ represents the deep feature of the test signal. The anomaly score $\mathrm{a}^{\text{maha}}$ can be computed as follows:
\begin{equation}
\label{eq:mahalanobis}
\mathrm{a}^{\text{maha}} = \sqrt{(\mathbf{x} - \boldsymbol{\mu}_{t,i})^\mathrm{T} (\boldsymbol{\Sigma}_{t,i})^{-1} (\mathbf{x} - \boldsymbol{\mu}_{t,i})}
\end{equation}

To determine the final anomaly score for a specific machine, we consider the scores from all three distinct perspectives, including the likelihood output $\mathrm{a^{\text{out}}}$ from $\mathcal{C}_{\text{type}}$, the probability output $\mathrm{a^{\text{arc}}}$ from $\mathcal{C}_{\text{id}}$, and the Mahalanobis distance $\mathrm{a}^{\text{maha}}$ from the latent space of the normal data. In order to combine the scores from various sources, we transform them into a standardized scale the same way described in~\cite{giri2020groupMADE}. First, for each kind of the three anomaly scores, mean and standard deviation are calculated over the training data for each machine ID. Then, each kind of anomaly score are standardized to have zero means and unit variance. Finally, with the standardized scores, we select the best combination of the scores from the training data to ensure accuracy. In this way, our evaluator can produce a comprehensive anomaly score that takes into account multiple factors, allowing us to effectively detect anomalies in various machines.

\begin{table*}[!htbp]
\centering
\setlength{\tabcolsep}{2.3pt}
\fontsize{8}{11}\selectfont
\caption{AUC {[}\%{]} for each machine type in MIMII dataset.}\vspace{-2mm}
\label{tab:mimii}
\begin{tabular}{@{}c|cccccccccccc|ccc|c@{}}
\toprule
\multirow{2}{*}{Methods} & \multicolumn{3}{c}{Fan}                          & \multicolumn{3}{c}{Pump}                         & \multicolumn{3}{c}{Valve}                        & \multicolumn{3}{c|}{Slide Rail}                  & \multirow{2}{*}{\begin{tabular}[c]{@{}c@{}}-6dB\\ Avg.\end{tabular}} & \multirow{2}{*}{\begin{tabular}[c]{@{}c@{}}0dB\\ Avg.\end{tabular}} & \multirow{2}{*}{\begin{tabular}[c]{@{}c@{}}6dB\\ Avg.\end{tabular}} & \multirow{2}{*}{\begin{tabular}[c]{@{}c@{}}Total\\ Avg.\end{tabular}} \\ \cmidrule(lr){2-13}
                         & -6dB           & 0dB            & 6dB            & -6dB           & 0dB            & 6dB            & -6dB           & 0dB            & 6dB            & -6dB           & 0dB            & 6dB            &                                                                      &                                                                     &                                                                     &                                                                       \\ \midrule
AE               & 68.73          & 84.85          & 95.30          & 71.00          & 81.61          & 86.86          & 50.26          & 54.86          & 59.47          & 73.42          & 78.49          & 90.27          & 65.85                                                                & 74.95                                                               & 82.98                                                               & 74.59                                                                 \\
Variational AE           & 71.47          & 84.88          & 94.76          & 70.97          & 81.68          & 87.69          & 49.79          & 54.67          & 57.29          & 70.54          & 78.11          & 89.74          & 65.69                                                                & 74.84                                                               & 82.37                                                               & 74.30                                                                 \\
GRLNet~\cite{sha2022GRLNet}      
                         & 69.93          & 86.62          & 95.34          & 77.46          & 85.31          & 90.12          & 53.41          & 57.01          & 63.93          & 74.97          & 80.85          & 91.10          & 68.94                                                                & 77.45                                                               & 85.12                                                               & 77.17                                                                 \\ \midrule
Score $\mathrm{a}^{\text{maha}}$          & 87.34          & 92.60          & 94.76          & 88.85          & 90.90          & 98.29          & 97.87          & 98.07          & 99.60          & 92.99          & 97.17          & 99.53          & 91.76                                                                & 94.69                                                               & 98.05                                                               & 94.83                                                                 \\
Score $\mathrm{a}^{\text{out}}$             & 83.16          & 88.41          & 80.85          & 88.22          & 89.37          & 97.25          & 88.90          & 94.03          & 95.88          & 85.42          & 95.80          & 98.59          & 86.43                                                                & 91.90                                                               & 93.14                                                               & 90.49                                                                 \\
Score $\mathrm{a}^{\text{arc}}$                 & 65.51          & 78.11          & 76.14          & 80.09          & 71.82          & 70.83          & 96.05          & 81.84          & 70.24          & 94.41          & 86.74          & 73.62          & 84.02                                                                & 79.63                                                               & 72.71                                                               & 78.78                                                                 \\
Combined Score       & \textbf{87.78} & \textbf{92.60} & \textbf{96.82} & \textbf{89.77} & \textbf{91.15} & \textbf{98.91} & \textbf{97.94} & \textbf{98.30} & \textbf{99.78} & \textbf{94.41} & \textbf{97.50} & \textbf{99.74} & \textbf{92.48}                                                       & \textbf{94.89}                                                      & \textbf{98.81}                                                      & \textbf{95.39}                                                        
                                                                 \\ \bottomrule
\end{tabular}%
\vspace{-3mm}
\end{table*}

\section{Experiments}
\vspace{-1mm}\subsection{Datasets}
To show the performance of our framework, we conduct experiments on two popular datasets: MIMII Dataset~\cite{purohit2019mimii} and DCASE~2020 Challenge Task~2 Dataset~\cite{koizumi2020description}.
\label{subsec:dataset}
\vspace{-2mm}\subsubsection{MIMII Dataset}
MIMII is a dataset that contains real-world industrial recordings for detecting anomalous machines. It contains 10-second 16-kHz recordings, recorded from four different machine types: fan, pump, valve and slide rail. Each type of machine contains four machine IDs. More importantly, the audio clips are augmented with three different signal-to-noise ratio~(SNR) to mimic the real-world industrial situation. In our experiment, we adopt the same train-test-split as shown in~\cite{sha2022GRLNet}.
\vspace{-2mm}\subsubsection{DCASE~2020 Challenge Task~2 Dataset}
DCASE 2020 Challenge Task~2 dataset contains 10-second single-channel 16-kHz recordings, selected from two datasets: MIMII~\cite{purohit2019mimii} and ToyADMOS~\cite{koizumi2019toyadmos}. Since we are interested in detecting malfunctions in industrial settings, we discard ToyADMOS part and only focus on MIMII part of the dataset, which contains seven machine IDs for each machine types.

\vspace{-1mm}\subsection{Implementation}
\label{subsec:implementation}
In this work, we employ a log-Mel spectrogram with 128 Mel filters as input, with the number of FFT points and hop length set to 1024 and 512 respectively. Our encoder contains three conformer blocks without positional encoding, with 512 linear units for FFN modules in each block. We adopt four heads in the MHSA module, with an output dimension of 128. To extract deep features, we utilize an attentive statistical pooling layer after the conformer blocks to get 64-dimensional features. In the ArcFace loss, we set the radius and margin to 16 and 1.28 respectively, with the intention of making the classifier more difficult to train and encouraging the encoder to learn better features. 

For the training strategy, we first freeze $\mathcal{C}_{\text{type}}$ to train the encoder for 80 epochs. Then, we unfreeze $\mathcal{C}_{\text{type}}$ and train for another 40 epochs. We use the ADAM optimizer~\cite{kingma20153rd} with learning rate equal to 0.001. The batch size is set to 28. Within each batch, we make sure that the number of samples of different machine IDs are the same, and we keep the same amount of total normal samples and pseudo-anomalous samples.

\vspace{-1mm}\subsection{Results}
\label{subsec:results}
We evaluate the ASD performance by calculating the area under the receiver operating characteristic curve~(AUC). To show the effectiveness of our model, we include the results of some competing systems for comparison.

\begin{table}[!htbp]
\vspace{-2mm}
\centering
\setlength{\tabcolsep}{3pt}
\fontsize{8}{11}\selectfont
\caption{AUC {[}\%{]} results for MIMII part in the DCASE 2020 development dataset.}\vspace{-2mm}
\label{tab:dc20_dev}
\begin{tabular}{@{}c|cccc|c@{}}
\toprule
Methods & Fan      & Pump      & Valve      & Slider    & Average     \\ \midrule
Official Baseline~\cite{koizumi2020description}           
                         & 65.83               & 72.89               & 66.28              & 84.76         & 72.44        \\
IDNN~\cite{suefusa2020IDNN}          
                         & 67.71                & 73.76                 & 84.09             & 86.45            & 78.00              \\
Glow\_Aff~\cite{dohi2021flow}     
                         & 74.90                   & 83.40              & 91.40            & 94.60           & 86.08          \\
GroupMADE~\cite{giri2020groupMADE}             
                         & 70.10               & 75.68            & 89.68           & 93.29           & 82.19           \\
IDCAE~\cite{daniluk2020IDCAE}                 
                         & 79.29            & 84.58             & 82.21               & 81.25           & 81.83                   \\ \midrule
Proposed Method   & \textbf{88.80} & \textbf{94.12}  & \textbf{100.00}  & \textbf{96.52} & \textbf{94.86}  \\ \bottomrule
\end{tabular}%
\vspace{-3mm}
\end{table}

Table~\ref{tab:mimii} presents the AUC results of our proposed approach on the MIMII dataset. Other scores mentioned in~\cite{sha2022GRLNet} are introduced for comparison. Our method achieves superior performance compared to the state-of-the-art method for all four machine types under varying SNR conditions. In the table, we report individual scores for all aspects in our approach~($\mathrm{a}^{\text{out}}$, $\mathrm{a}^{\text{arc}}$, and $\mathrm{a}^{\text{maha}}$). Among these scores, $\mathrm{a}^{\text{maha}}$ yields the highest score and contributes the most to the overall performance, indicating that our approach effectively learns inlier distribution of the normal data. Moreover, the overall score is superior to all the individual scores, demonstrating that the ASD performance can be enhanced by considering both inlier and outlying factors.

\begin{table}[!htbp]
    \centering
    \setlength{\tabcolsep}{1pt}
    \fontsize{8}{11}\selectfont
    \caption{AUC {[}\%{]} results for MIMII part in the DCASE 2020 evaluation dataset and System Complexity~(Comp.).}\vspace{-2mm}
    \label{tab:dc20_eval}
    \begin{threeparttable}
        \begin{tabular}{@{}c|cccc|c|c@{}}
            \toprule
            Methods & Fan         & Pump        & Valve     & Slider     & Average  & Comp. \\ \midrule
            Official Baseline~\cite{koizumi2020description}        
                                     & 82.80             & 82.37         & 57.37             & 79.41                  & 75.49 & 269K \\
            GroupMADE~\cite{giri2020groupMADE}             
                                     & 84.52              & 88.07              & 84.23             & 95.18           & 88.00 & 663K  \\
            IDCAE~\cite{daniluk2020IDCAE}                 
                                     & 90.70               & \textbf{92.65} & 88.01             & 88.01              & 89.84  & 2M  \\ 
            DDCSAD~\cite{kuroyanagi2021anomalous}
                                     & 95.14               & 92.14  & 96.08  & 97.60 & 95.24  & 5M$^\ast$   \\ \midrule
            Proposed Method   & \textbf{95.42} & 91.72& \textbf{97.33}  & \textbf{97.69}& \textbf{95.54} & 2M \\ \midrule
            Fused System in~\cite{giri2020groupMADE}$^{\dag}$ & 94.54 & 93.65 & 96.13 & 97.63 & 95.49 & 2M \\
            Fused System in~\cite{daniluk2020IDCAE}$^{\ddag}$  & 99.13 & 95.07 & 90.97 & 98.18 & 95.84 & 179M \\ \bottomrule
        \end{tabular}%
        \begin{tablenotes}
            \footnotesize
            \item[$\ast$] {\small The authors did not reveal the model in~\cite{kuroyanagi2021anomalous}. This is our estimation based on their model in~\cite{kuroyanagi2022improvement}.}
            \item[\dag] {\small DCASE 2020 Task 2 1st-ranked 2-system fusion.}
            \item[\ddag] {\small DCASE 2020 Task 2 2nd-ranked 3-system fusion.}
        \end{tablenotes}
    \end{threeparttable}
\vspace{-6mm}
\end{table}


Table~\ref{tab:dc20_dev} and Table~\ref{tab:dc20_eval} present the AUC results on the DCASE 2020 Challenge dataset. To show the superiority of our method, we include the scores from several competing single-model systems that uses only one model in the front-end in their framework. As it is depicted in Table~\ref{tab:dc20_dev}, our approach achieves best performance on all types of machines on the development dataset. We also test our approach on the evaluation dataset using the best training model, and the results are shown in Table~\ref{tab:dc20_eval}. 
We outperform the baseline model by a large margin, and achieve best performance comparing to the state-of-the-art single-model systems on all four machine types except for pump. In the table, we also list the results from the systems of the top~2 teams~\cite{giri2020groupMADE,daniluk2020IDCAE} in the challenge for comparison. Our method with a single-model structure has a comparable ASD performance with the multi-system fusion ones, but with lower system complexity. This indicates that with only one model, we can achieve the similar goal that used to be done by IM-OE ensembles.

\vspace{-3mm}\section{Conclusions}
In this paper, we propose a novel approach for ASD that uses MTL to teach a conformer-based encoder to learn effective features by outlier-aware inlier modeling. In the inference phase, our approach can provide anomaly scores from multiple perspectives while using only one model, making it a simpler and more efficient alternative to ensemble systems. The experimental results on multiple datasets show that our approach outperforms other state-of-the-art single-model systems and achieves comparable performance compared to the top-ranked ensemble systems. In future work, we plan to investigate the potential of using single-model MTL framework for detecting anomalies across different domains.

\vspace{0mm}\section{Acknowledgements}
This research is funded in part by the Science and Technology Program of Suzhou City~(SYC2022051), National Natural Science Foundation of China~(62171207) and OPPO. Many thanks for the computational resource provided by the Advanced Computing East China Sub-Center.

\vfill\pagebreak

\bibliographystyle{IEEEtran}
\bibliography{mybib}

\begin{thebibliography}{10}
\providecommand{\url}[1]{#1}
\csname url@samestyle\endcsname
\providecommand{\newblock}{\relax}
\providecommand{\bibinfo}[2]{#2}
\providecommand{\BIBentrySTDinterwordspacing}{\spaceskip=0pt\relax}
\providecommand{\BIBentryALTinterwordstretchfactor}{4}
\providecommand{\BIBentryALTinterwordspacing}{\spaceskip=\fontdimen2\font plus
\BIBentryALTinterwordstretchfactor\fontdimen3\font minus
  \fontdimen4\font\relax}
\providecommand{\BIBforeignlanguage}[2]{{%
\expandafter\ifx\csname l@#1\endcsname\relax
\typeout{** WARNING: IEEEtran.bst: No hyphenation pattern has been}%
\typeout{** loaded for the language `#1'. Using the pattern for}%
\typeout{** the default language instead.}%
\else
\language=\csname l@#1\endcsname
\fi
#2}}
\providecommand{\BIBdecl}{\relax}
\BIBdecl

\bibitem{kamat2020anomaly}
P.~Kamat and R.~S. Dr., ``Anomaly detection for predictive maintenance in
  industry 4.0- a survey,'' \emph{E3S Web of Conferences}, 2020.

\bibitem{tanuska2021smart}
P.~Tanuska, L.~Spendla, M.~Kebisek, R.~Duris, and M.~Stremy, ``Smart anomaly
  detection and prediction for assembly process maintenance in compliance with
  industry 4.0,'' \emph{Sensors}, vol.~21, no.~7, p. 2376, 2021.

\bibitem{wu2021network}
H.~Wu, Y.~Shen, X.~Xiao, A.~Hecker, and F.~H. Fitzek, ``In-network processing
  acoustic data for anomaly detection in smart factory,'' in \emph{Proc. of
  GLOBECOM}, 2021, pp. 1--6.

\bibitem{koizumi2020description}
Y.~Koizumi, Y.~Kawaguchi, K.~Imoto, T.~Nakamura, Y.~Nikaido, R.~Tanabe,
  H.~Purohit, K.~Suefusa, T.~Endo, M.~Yasuda, and N.~Harada, ``Description and
  discussion on dcase2020 challenge task2: Unsupervised anomalous sound
  detection for machine condition monitoring,'' \emph{ArXiv}, vol.
  abs/2006.05822, 2020.

\bibitem{kawaguchi2021description}
Y.~Kawaguchi, K.~Imoto, Y.~Koizumi, N.~Harada, D.~Niizumi, K.~Dohi, R.~Tanabe,
  H.~Purohit, and T.~Endo, ``Description and discussion on dcase 2021 challenge
  task 2: Unsupervised anomalous sound detection for machine condition
  monitoring under domain shifted conditions,'' \emph{ArXiv}, vol.
  abs/2106.04492, 2021.

\bibitem{chandola2009anomaly}
V.~Chandola, A.~Banerjee, and V.~Kumar, ``Anomaly detection: A survey,''
  \emph{ACM Comput. Surv.}, vol.~41, pp. 15:1--15:58, 2009.

\bibitem{giri2020groupMADE}
R.~Giri, S.~V. Tenneti, K.~Helwani, F.~Cheng, U.~Isik, and A.~Krishnaswamy,
  ``Unsupervised anomalous sound detection using self-supervised classification
  and group masked autoencoder for density estimation,'' \emph{DCASE 2020
  Challenge, Tech. Rep}, vol.~23, 2020.

\bibitem{daniluk2020IDCAE}
P.~Daniluk, M.~Go{\'z}dziewski, S.~Kapka, and M.~Ko{\'s}mider, ``Ensemble of
  auto-encoder based and wavenet like systems for unsupervised anomaly
  detection,'' \emph{DCASE 2020 Challenge, Tech. Rep}, 2020.

\bibitem{hayashi2020conformer}
T.~Hayashi, T.~Yoshimura, and Y.~Adachi, ``Conformer-based id-aware autoencoder
  for unsupervised anomalous sound detection,'' \emph{DCASE 2020 Challenge,
  Tech. Rep.}, 2020.

\bibitem{suefusa2020IDNN}
K.~Suefusa, T.~Nishida, H.~Purohit, R.~Tanabe, T.~Endo, and Y.~Kawaguchi,
  ``Anomalous sound detection based on interpolation deep neural network,'' in
  \emph{Proc. of ICASSP}, 2020, pp. 271--275.

\bibitem{breunig2000lof}
M.~M. Breunig, H.-P. Kriegel, R.~T. Ng, and J.~Sander, ``Lof: identifying
  density-based local outliers,'' in \emph{Proc. of ACM SIGMOD}, 2000, pp.
  93--104.

\bibitem{scott2004outlier}
D.~W. Scott, ``Outlier detection and clustering by partial mixture modeling,''
  in \emph{Proc. of COMPSTAT}, 2004, pp. 453--464.

\bibitem{liu2019outlier}
W.~Liu, D.~Cui, Z.~Peng, and J.~Zhong, ``Outlier detection algorithm based on
  gaussian mixture model,'' in \emph{Proc. of ICPICS}, 2019, pp. 488--492.

\bibitem{dohi2021flow}
K.~Dohi, T.~Endo, H.~Purohit, R.~Tanabe, and Y.~Kawaguchi, ``Flow-based
  self-supervised density estimation for anomalous sound detection,'' in
  \emph{Proc. of ICASSP}, 2021, pp. 336--340.

\bibitem{kuroyanagi2022improvement}
I.~Kuroyanagi, T.~Hayashi, K.~Takeda, and T.~Toda, ``Improvement of serial
  approach to anomalous sound detection by incorporating two binary
  cross-entropies for outlier exposure,'' in \emph{Proc. of EUSIPCO}, 2022, pp.
  294--298.

\bibitem{primus2020reframing}
P.~Primus, ``Reframing unsupervised machine condition monitoring as a
  supervised classification task with outlier-exposed classifiers,''
  \emph{DCASE 2020 Challenge, Tech. Rep.}, 2020.

\bibitem{inoue2020detection}
P.~Vinayavekhin, T.~Inoue, S.~Morikuni, S.~Wang, T.~H. Trong, D.~Wood,
  M.~Tatsubori, and R.~Tachibana, ``Detection of anomalous sounds for machine
  condition monitoring using classification confidence,'' \emph{DCASE 2020
  Challenge, Tech. Rep.}, 2020.

\bibitem{lopez2021ensemble}
J.~A. Lopez, G.~Stemmer, P.~Lopez-Meyer, P.~Singh, J.~A. del Hoyo~Ontiveros,
  and H.~A. Cordourier, ``Ensemble of complementary anomaly detectors under
  domain shifted conditions.'' in \emph{DCASE}, 2021, pp. 11--15.

\bibitem{morita2021anomalous}
K.~Morita, T.~Yano, and K.~Tran, ``Anomalous sound detection using cnn-based
  features by self supervised learning,'' \emph{DCASE 2021 Challenge, Tech.
  Rep.}, 2021.

\bibitem{wilkinghoff2021utilizing}
K.~Wilkinghoff, ``Utilizing sub-cluster adacos for anomalous sound detection
  under domain shifted conditions,'' \emph{DCASE 2021 Challenge, Tech. Rep.},
  2021.

\bibitem{georgescu2021anomaly}
M.-I. Georgescu, A.~Barbalau, R.~T. Ionescu, F.~S. Khan, M.~Popescu, and
  M.~Shah, ``Anomaly detection in video via self-supervised and multi-task
  learning,'' in \emph{Proc. of CVPR}, 2021, pp. 12\,742--12\,752.

\bibitem{doshi2022multi}
K.~Doshi and Y.~Yilmaz, ``Multi-task learning for video surveillance with
  limited data,'' in \emph{Proc. of CVPR}, 2022, pp. 3889--3899.

\bibitem{kuroyanagi2021anomalous}
I.~Kuroyanagi, T.~Hayashi, K.~Takeda, and T.~Toda, ``Anomalous sound detection
  using a binary classification model and class centroids,'' in \emph{Proc. of
  EUSIPCO}, 2021, pp. 1995--1999.

\bibitem{gulati2020conformer}
A.~Gulati, J.~Qin, C.-C. Chiu, N.~Parmar, Y.~Zhang, J.~Yu, W.~Han, S.~Wang,
  Z.~Zhang, Y.~Wu \emph{et~al.}, ``Conformer: Convolution-augmented transformer
  for speech recognition,'' \emph{Proc. of Interspeech}, pp. 5036--5040, 2020.

\bibitem{deng2019arcface}
J.~Deng, J.~Guo, N.~Xue, and S.~Zafeiriou, ``Arcface: Additive angular margin
  loss for deep face recognition,'' in \emph{Proc. of CVPR}, 2019, pp.
  4690--4699.

\bibitem{hojjati2022self}
H.~Hojjati and N.~Armanfard, ``Self-supervised acoustic anomaly detection via
  contrastive learning,'' in \emph{Proc. of ICASSP}, 2022, pp. 3253--3257.

\bibitem{sha2022GRLNet}
Y.~Sha, S.~Gou, J.~Faber, B.~Liu, W.~Li, S.~Schramm, H.~Stoecker,
  T.~Steckenreiter, D.~Vnucec, N.~Wetzstein \emph{et~al.}, ``Regional-local
  adversarially learned one-class classifier anomalous sound detection in
  global long-term space,'' in \emph{Proc. of ACM SIGKDD}, 2022, pp.
  3858--3868.

\bibitem{purohit2019mimii}
H.~Purohit, R.~Tanabe, K.~Ichige, T.~Endo, Y.~Nikaido, K.~Suefusa, and
  Y.~Kawaguchi, ``Mimii dataset: Sound dataset for malfunctioning industrial
  machine investigation and inspection,'' \emph{ArXiv}, vol. abs/1909.09347,
  2019.

\bibitem{koizumi2019toyadmos}
Y.~Koizumi, S.~Saito, H.~Uematsu, N.~Harada, and K.~Imoto, ``Toyadmos: A
  dataset of miniature-machine operating sounds for anomalous sound
  detection,'' in \emph{Proc. of WASPAA}, 2019, pp. 313--317.

\bibitem{kingma20153rd}
D.~P. Kingma and J.~Ba, ``Adam: A method for stochastic optimization,''
  \emph{CoRR}, vol. abs/1412.6980, 2014.

\end{thebibliography}

\end{document}